\documentclass[twocolumn,aps,prd,nofootinbib]{revtex4-1}

\pdfoutput=1

\usepackage{amsmath, amsfonts,amsthm, amssymb, mathrsfs, bm, verbatim, pslatex, bigints, physics, tikz, xcolor, graphicx,tikz-feynman}
\usetikzlibrary{decorations.markings}

\newcommand{\be}{\begin{equation}}
\newcommand{\ee}{\end{equation}}
\newcommand{\bea}{\begin{eqnarray}}
\newcommand{\eea}{\end{eqnarray}}

\newcommand{\s}{j}
\newcommand{\g}{G}
\newcommand{\m}{\eta}
\newcommand{\p}{\pi}
\newcommand{\kk}{K}
\newcommand{\ff}{F}
\newcommand{\wf}{\psi}
\newcommand{\K}{M}
\newcommand{\ca}{a}
\newcommand{\cb}{b}
\newcommand{\cc}{c}
\newcommand{\uv}{\epsilon}
\newcommand{\Kr}{J}

\newcommand{\limV}{\lim_{V\to\infty}}
\newcommand{\limR}{\lim_{R\to\infty}}
\newcommand{\limS}{\lim_{S\to\infty}}
\newcommand{\V}{\bar{V}}

\begin{document}

\title{Ground State Wave Function Overlap in Superconductors and Superfluids}

\author{Mark P. Hertzberg}
\email{mark.hertzberg@tufts.edu}
\affiliation{Institute of Cosmology, Department of Physics and Astronomy,
Tufts University, Medford, MA 02155, USA}

\author{Mudit Jain}
\email{mudit.jain@tufts.edu}
\affiliation{Institute of Cosmology, Department of Physics and Astronomy,
Tufts University, Medford, MA 02155, USA}

\begin{abstract}
In order to elucidate the quantum ground state structure of non-relativistic condensates, we explicitly construct the ground state wave function for multiple species of bosons, describing either superconductivity or superfluidity. Since each field $\Psi_j$ carries a phase $\theta_j$ and the Lagrangian is invariant under rotations $\theta_j\to\theta_j+\alpha_j$ for independent $\alpha_j$, one can investigate the corresponding wave function overlap between a pair of ground states $\langle G|G'\rangle$ differing by these phases. We operate in the infinite volume limit and use a particular prescription to define these states by utilizing the position space kernel and regulating the UV modes. We show that this overlap vanishes for most pairs of rotations, including $\theta_j\to\theta_j+m_j\,\epsilon$, where $m_j$ is the mass of each species, while it is unchanged under the transformation $\theta_j\to\theta_j+q_j\,\epsilon$, where $q_j$ is the charge of each species. We explain that this is consistent with the distinction between a superfluid, in which there is a non-trivial conserved number, and the superconductor, in which the electric field and conserved charge is screened, while it is compatible with a non-zero order parameter in both cases. Moreover, we find that this bulk ground state wave function overlap directly reflects the Goldstone boson structure of the effective theory and provides a useful diagnostic of its physical phase.
\end{abstract}

\maketitle

\section{Introduction}

The structure of the ground state of a field theory is a highly important subject, with applications to the Standard Model, cosmology, and condensed matter systems. As an example, in the context of the Standard Model, there are various forms of (approximate) spontaneously broken symmetries (SSB) in the vacuum, including the breakdown of chiral symmetry in QCD, etc. However, the structure of the vacuum within the Higgs mechanism often involves some confusion, since it is often described as being tied to the notion of a a breakdown of gauge symmetry, which is in fact a type of redundancy in the description. This has led to various conclusions in the literature; see Refs.~\cite{Englert:1964et,Higgs:1964pj,Guralnik:1964eu,Bernstein:1974rd,Strocchi:1977za,Stoll:1995yg,Maas:2012ct,Kibble:2014gug,Strocchi:2015uaa,Maas:2017wzi}. 

To elucidate the structure of the vacuum in the Standard Model, we recently constructed the (approximate) vacuum wave function of the Standard Model and explicitly found the wave function overlap between vacua \cite{Hertzberg:2018kyi}. There is other important older work, including the famous Elitzur's theorem \cite{Elitzur:1975im} -- that one cannot spontaneously break a local, or ``small", gauge symmetry (which is obvious since they are always only redundancies).

In this work, we turn our attention to nonrelativistic condensed matter systems. In particular we will focus on multi-fluid systems of bosons. Some of the most familiar applications are to collections of helium atoms, which can organize into a superfluid at low temperatures, and to collections of Cooper pairs of electrons, which can organize into a superconductor at low temperatures. The general topic of superfluids and superconductors will be the subject of this paper (for some foundations and reviews, see Refs.~\cite{Landau,Ginzburg,Bardeen:1957mv,Schriefferbook,Feynman,London,Leggett:1975te,Wheatley:1975tf,Rabinowitz,Guenault,Annett,Schmitt:2014eka,Sharma,Stewart}). 

In the context of superfluidity, it is well known that there is a global $U(1)$ phase rotation symmetry of the Schrodinger field $\psi$ that is spontaneously broken by the ground state; its corresponding Goldstone is a phonon associated with gapless sound waves. In the context of superconductivity, it is well known that the system can exhibit plasma oscillations; and there is a gapped spectrum provided by the plasma frequency. For a discussion of Goldstones in condensed matter systems, see Refs.~\cite{Brauner:2010wm,Watanabe:2012hr,Nicolis:2012vf,Watanabe:2014fva}.

This absence of a Goldstone mode in the latter case, has led to various contradictory statements in the literature surrounding the ground state of a superconductor.\footnote{For example, in the textbook ``The quantum theory of fields" \cite{WeinbergQFT} it is claimed ``{\em A superconductor is simply a material in which electromagnetic gauge invariance is spontaneously broken}." However, one cannot spontaneously break gauge invariance, as it is a mere redundancy, as mentioned above, and all states are gauge invariant. However, one can wonder about the fate of the {\em global} sub-group of $U(1)_{em}$, which after all is the actual symmetry of electromagnetism, associated with conservation of electric charge. Is it possible that this symmetry is spontaneously broken in superconductors, despite the absence of the gapless Goldstone mode? In the review paper \cite{Greiter}, it was claimed that this is precisely what happens ``{\em global $U(1)$ phase rotation symmetry, and not gauge symmetry, is spontaneously violated}". But how could there possibly be SSB when there is no associated Goldstone mode due to plasma oscillations? Ref.~\cite{Henley} claimed ``{\em in a superconductor the superfluid density fluctuations carry charge density fluctuations, which have long-range Coulomb interactions, whereas Goldstone's theorem only applies to local interactions}". Thus claiming the Goldstone theorem is avoided and there is still SSB. However, since the electric field gets screened this reasoning is unclear. 
On the other hand, in another work Ref.~\cite{Kovner} claimed that the SSB pattern is just the opposite of this ``{\em in the superconducting phase the symmetry is unbroken}" (which accurately reflects the presence of otherwise of massless modes in the two phases). In Ref.~\cite{Poniatowski} the symmetry breaking language is said to be just ``{\em linguistics}" and that ``{\em there is only one physical ground state}". And Ref.~\cite{Nima} refers to the (relativistic) Higgs mechanism as ``{\em nothing to do with ‘symmetry breaking}’" as ``{\em there are even gauges (such as unitary gauge) where there is no ‘symmetry breaking’ description of the physics}."}
In this paper our goal is to address a very specific aspect of the structure of the ground state in a direct and clear fashion, building on the ideas we developed in Ref.~\cite{Hertzberg:2018kyi}. In particular, we will explicitly compute the ground state wave function $|\g\rangle$ in a particular well defined prescription, and then perform the (global) phase rotation transformations $\theta_\s\to\theta_\s+\alpha_\s$ to obtain other possible ground states $|\g'\rangle$. We then explicitly compute the overlap $\langle \g|\g'\rangle$ to determine which, if any, phase rotations lead to new states. This will be the particular aspect of the ground state structure which we will focus on. We will make use of a very specific prescription to define the ground state, as it is subtle in the infinite volume limit. Importantly, we will use a prescription that treats the superfluid and superconducting cases altogether in the same computation. Of course, there are other important aspects of the ground state, including other issues surrounding SSB. This includes the existence of non-zero expectation values of certain kinds of field operators which can provide novel order parameters; however, these other issues have been very well studied elsewhere and are not the focus of the present study. 

To our knowledge the explicit construction of the wave function overlap in this context does not appear directly in the literature. Our focus here will be on the properties within the bulk of the material; we will then comment on effects from the boundary in the conclusions. By defining the ground state through a kernel representation in position space and carefully regulating the UV modes, we will find a beautiful connection between the structure of the overlap $\langle \g|\g'\rangle$ and the Goldstone boson pattern, which is the primary finding of this work.

The class of models we will study is the effective theory of bosons (since Cooper pairs organize into bosons in the condensed phase) in the nonrelativistic approximation (see Appendix \ref{RA} for leading relativistic corrections). For generality we will study a multi-component system, specializing to a two-fluid system, as this will allow us to identify more details of the ground state structure, as we will explain. Each bosonic field is described by a Schrodinger field $\Psi_\s$ ($\s=1,2$), with phase $\theta_\s$ formally displaying a pair of phase rotation symmetries $\theta_\s\to\theta_\s+\alpha_\s$, for arbitrary choices of $\alpha_\s$. We allow for a coupling to electromagnetism by endowing each species with charge $q_\s$ (but in such a way that the homogeneous background charge density is zero). We find that the overlap of the wave functions is zero for most choices of $\alpha_\s$, including $\alpha_\s=m_\s\,\epsilon$ where $m_\s$ is the mass of each of the species. This implies SSB and is associated with the conservation of mass. On the other hand, we find that for the special choice of $\alpha_\s=q_\s\,\epsilon$, the overlap of the wave functions is 1. 
This makes the overlap wave function we construct to be directly associated with the Goldstone boson structure and a useful diagnostic of the physical phase that the system is in.

Let us note that in this broader class of models, including multiple species of bosons, in addition to the electric symmetry, it also possesses other global symmetries. There is no reason a priori to expect that those symmetries cannot be spontaneously broken in the usual way. The electric symmetry on the other hand, since it does not have a local order parameter, is different and therefore we expect its realization to be distinct. It is useful to see exactly how these different properties are realized on the ground state wave function; this is non obvious because these various transformations $\theta_\s\to\theta_\s+\alpha_\s$ are mixed up with one another, and are coupled because the electric interaction couples the multiples species to one another. The idea that the electric symmetry is expected to be realized differently is for a very physical reason that we shall return to: the conserved charge $Q$ in the superconductor case behaves differently to the conserved number $N$ of a regular global symmetry in the superfluid case. The reason being that in a superconductor the electric field is screened, so that when one integrates the electric flux over a surface at infinity, one formally obtains a vanishing charge; while the conserved number of a superfluid does not have any analogous screening. Since conserved quantities generate symmetries, this again leads to the expectation that the structure of the ground state is distinct. However, we will simply compute the overlap at infinite volume, using the same prescription in both superfluid and superconducting cases, and see the outcome directly.

Our paper is organized as follows:
In Section \ref{Nonrelativistic} we present the two-fluid nonrelativistic model.
In Section \ref{Homogeneous} we discuss the condensate background.
In Section \ref{Perturbations} we present the Lagrangian and Hamiltonian governing fluctuations.
In Section \ref{Ground} we explicitly compute the ground state wave function.
In Section \ref{Overlap} we explicitly compute the overlap of the ground state wave functions.
In Section \ref{Conserved} we discuss which quantities are conserved and the associated symmetries.
Finally, in Section \ref{Conclusions} we conclude.

\section{Nonrelativistic Field Theory}\label{Nonrelativistic}

We are interested in systems of nonrelativistic bosons. We will allow for several species that are distinguishable, labelled $\s$, but will often specialize to the case of two species ($\s=1,2$) where needed for simplicity. We are interested in systems of many particles. In this case, it is convenient to utilize creation $\hat{a}^\dagger_{{\bf k},j}$ and destruction operators $\hat{a}_{{\bf k},\s}$ that produce $N$ particle states. Since we are interested in exploring condensates, it is convenient to pass to the field representation. This is defined by Fourier transforming the destruction operator to a field in position space, the so-called Schrodinger field $\hat{\Psi}_\s({\bf x})$ (with conjugate field $\hat{\Psi}^\dagger_\s({\bf x})$) as follows
\be
\hat{\Psi}_\s({\bf x}) = \int\!\!{d^3k\over(2\pi)^3}\,\hat{a}_{{\bf k},\s}\,e^{i{\bf k}\cdot{\bf x}}.
\ee
The corresponding particle number density operator for species $\s$ is
\be
\hat{n}_\s({\bf x})= \hat{\Psi}^\dagger_\s({\bf x}) \,\hat{\Psi}_\s({\bf x}),
\ee
with particle number operator $\hat{N}_\s=\int d^3x\,\,\hat{n}_\s({\bf x})$.

In order to explore superconductivity, we minimally couple the fields to electromagnetism $A_\mu=(-\phi,{\bf A})$. For convenience we will use the Lagrangian formalism (though later we will move to the Hamiltonian formalism). 
Since ordinary superconductors and superfluids involve electrons and nuclei moving much slower than the speed of light, one can often use an effective nonrelativistic description. In order to build this, we note that the leading order scalar field sector is essentially specific uniquely by the Galilean symmetry. Furthermore, there is a unique way to couple to photons from the minimal coupling procedure. This uniquely specifies the nonrelativistic field theory. For the leading relativistic corrections, the reader may see  Appendix \ref{RA} for the sake of completeness. But for the most part, the nonrelativistic effective field theory will suffice, and is given by
\bea
\mathcal{L}=\sum_\s\left[{i\over2}\Psi_\s^*(\dot{\Psi}_\s + iq_\s \phi \Psi_\s)+c.c -{1\over 2m_\s}|\nabla\Psi_\s-i q_\s{\bf A}\psi_\s|^2\right]\nonumber\\
-V(\Psi)-{1\over4}F_{\mu\nu}F^{\mu\nu},\hspace{5cm}
\label{eqn:Lagrangian}
\eea
where $m_\s$ is the mass of each of the species and $q_\s$ is the charge of each of the species. For the potential $V$, we allow 4-point self-interactions. However, for simplicity we assume that each species does not directly scatter off other species. For instance, one can imagine that the underlying fermionic description, which introduces Pauli exclusion, gives rise to repulsion among the indistinguishable particles. The generalization to other couplings is straightforward, but will not be studied here. We also include a chemical potential $\mu_\s$ for each of the species, to make it simpler to describe a background (this can also be obtained from a redefinition of the fields as $\Psi_\s\to\Psi_\s\,e^{i\mu_\s t}$). Together we write the potential as
\be
V(\Psi)= \sum_\s\left[-\mu_\s|\Psi_\s|^2 + {\lambda_\s\over 2}|\Psi_\s|^4\right],
\ee
where $\lambda_\s$ are (positive) self-couplings. 

We note that the above theory carries the following set of (global) symmetries
\be
\Psi_\s\to \Psi_\s e^{i\alpha_\s},
\label{symmetry}\ee
for independent $\alpha_\s$. When expanding around the vacuum, these are associated with the conservation of each of the species particle number, and includes the special case of $\alpha_\s = q_\s\,\epsilon$ corresponding to electric charge. In the following we will examine their behavior for a condensate (ground state) solution.

\section{Homogeneous Background}\label{Homogeneous}

We now expand around a homogeneous background. At the classical field level, the ground state is determined by minimizing the above potential $V$. We can use the chemical potential to obtain whatever background number density of particles we desire. Lets denote the background number density of each species $n_{\s 0}$. By minimizing the potential, this value of number density can be immediately obtained by choosing the chemical potential to be
\be
\mu_\s = \lambda_j\,n_{\s 0}.
\ee
We will ensure that the background charge density $\rho_0=0$, so that we have a neutral superconductor to expand around. Hence we assume that the background number densities are related by the condition
\be
\rho_0=\sum_j q_\s\,n_{\s 0} +(q_n\, n_{n0}) = 0.
\ee
where we have included a possible $q_n n_{n0}$ term; this should be included in the case of a superconductor: it refers to nuclei, which carry positive charge $q_n>0$. The nuclei provide the compensating charge, so there is a well defined neutral superconductor, that we may then add charge density fluctuations to. This is the standard physical starting point. In the case of the superconductor, we will not need to track the dynamics of the nuclei (although they will inevitably play a role when talking about mass density perturbations), as they are not accurately described by bosons, and are very heavy; so they will only be relevant at this background level. Instead, as is well known, in a BCS superconductor, the relevant dynamics is provided by the much lighter Cooper pairs, with charge $q=-2 e$. One can have more general systems with multiple types of effective bosons. We will leave our analysis in terms of multiple species for the sake of pedagogy, and in fact the case of 2 species of bosons will often be our focus, since we can then discuss the fate of multiple types of symmetries. On the other hand, in the case of a superfluid, then the relevant species all carry no charge. In this case, we will just use the $j$ index to refer to the appropriate bosonic degrees of freedom, which are all heavy neutral bosons, such as helium, etc. Hence our framework is quite general.

The corresponding background field value for each of the relevant dynamical species (which is quite different depending on whether it is a superconductor or a superfluid) is given by
\be
v_\s\equiv |\Psi_{\s 0}|=\sqrt{n_{\s 0}}.
\ee
As is well known, the phase of the ground state condensate $\Psi_{\s 0}$ is not determined by this condition. This suggests there are a family of distinct ground state solutions labelled by a set of constant phase parameters $\theta_{\s 0}$ as
\be
\psi_{\s 0}=v_\s\,e^{i\theta_{\s 0}},
\ee
which all formally spontaneously break the symmetry given above in Eq.~(\ref{symmetry}). Since these symmetry transformations include the global sub-group of $U(1)_{em}$, one should be extra careful in aspects of this conclusion. In this work, we will examine the specific issue of the structure of the ground state systematically; first we quantize the fluctuations and actually computing the ground state of the quantum theory precisely.

\section{Perturbations}\label{Perturbations}

Let us expand around the homogeneous background by decomposing the fields into a perturbation in modulus $\eta_\s({\bf x},t)$ and phase parameter $\theta_\s({\bf x},t)$ as
\be
\Psi_\s({\bf x},t) = (v_\s + \m_\s({\bf x},t))e^{i\theta_\s({\bf x},t)}.
\label{eqn:perturbed.field}
\ee
We then treat $\m_\s$ and (derivatives of) $\theta_\s$ as small to study small perturbations. Expanding the Lagrangian density to quadratic order in the fluctuations we obtain
\bea
&&\mathcal{L}_2 = \sum_\s\Big{[}-2v_\s\m_\s\dot{\theta}_\s - 2v_\s q_\s\phi \,\m_\s -2\mu_\s\m_\s^2\nonumber\\
&&\,\,\,\,\,\,\,\,\,\,\,\,\,\,\,\,\,\,\,-{1\over2m_\s}((\nabla\m_\s)^2+v_\s^2(\nabla\theta_\s-q_\s{\bf A})^2)\Big{]}-{1\over4}F_{\mu\nu}F^{\mu\nu}.\,\,\,\,
\eea
Now the electromagnetic field includes the non-dynamical Coulomb potential $\phi$. We can solve for this from Gauss law as follows
\be
-\nabla^2\phi=\nabla\cdot\dot{\bf A}+\rho,
\ee
where the charge density (to linear order in perturbations) is
\be
\rho =\sum_\s 2 v_\s q_\s\m_\s.
\ee

We now decompose the vector potential ${\bf A}$ into its longitudinal ${\bf A}^L$ and transverse ${\bf A}^T$ components
\be
{\bf A} = {\bf A}^L+{\bf A}^T.
\ee 
However, we can now exploit gauge invariance to simplify our results by operating in Coulomb gauge $\nabla\cdot{\bf A}=0$. So ${\bf A}^L=0$ and ${\bf A}={\bf A}^T$ is purely transverse. One should bear in mind that all of our results can be trivially re-written in a gauge invariant way be replacing
\be
\theta_\s({\bf x})\to\theta_\s({\bf x})-q_\s{\nabla\cdot {\bf A}^L\over\nabla^2}
\ee
if desired. The Lagrangian density decomposes into a sum of longitudinal $\mathcal{L}_L$ and transverse $\mathcal{L}_T$ pieces that decouple at the quadratic order
\bea
&&\mathcal{L}_L = \sum_\s\Big{[}-2v_\s\m_\s\dot{\theta}_\s -2\mu_\s\m_\s^2-{1\over2m_\s}((\nabla\m_\s)^2+v_\s^2(\nabla\theta_\s)^2)\Big{]}\nonumber\\
&&\hspace{0.8cm}-2\left(\sum_\s{q_\s v_\s \nabla\m_\s\over\nabla^2}\right)^{\!2},\label{longitudinal}\\
&&\mathcal{L}_T = {1\over 2}(\dot{\bf A}^T)^2-{1\over2}(\nabla\times{\bf A}^T)^2-\sum_\s{v_\s^2q_\s^2\over 2m_\s}({\bf A}^T)^2.\label{transverse}
\eea
The final term in (\ref{transverse}) shows the familiar fact that within a superconductor the magnetic field acquires an effective mass. It is given by the sum of squares of the plasma frequencies as 
\be
m_{\tiny\mbox{eff}}^2 = \sum_\s{v_\s^2q_\s^2\over m_\s}.
\ee
This means the magnetic field is short ranged, which is the famous Meissner effect (for example, see \cite{Wentzel,Essen,Hirsch}). This is analogous to the Higgs mechanism in the Standard Model. However there is an important difference: In the Standard Model the Lorentz symmetry ensures that the Coulomb potential $A_0=-\phi$ also acquires the same effective mass. However, in this nonrelativistic setup that is not the case. As we will later discuss, despite appearances, the Coulomb potential (in Coulomb gauge) and the associated electric field gets screened more strongly than the magnetic field. This has important ramifications for the behavior of the charge and the fate of symmetries, as we will discuss in Section \ref{Conserved}.

Our interest is in the behavior of the longitudinal modes, as these involves the phase parameters $\theta_\s$, and enjoy the symmetries $\theta_\s\to\theta_\s+\alpha_\s$. To study these modes in more detail, it is convenient to now pass to the Hamiltonian formalism. The appropriate phase space variables are the phase $\theta_\s$ and momentum conjugates $\p_\s$ given by
\be
\p_\s={\partial\mathcal{L}\over\partial\dot{\theta}_\s}=-2v_\s\m_\s.
\ee
Furthermore, we diagonalize the problem by passing to $k$-space. We write the Hamiltonian for the longitudinal modes as
\be
H^L= \int\! {d^3k\over(2\pi)^3}\,\mathcal{H}^L_k
\ee
and find the $k$-space Hamiltonian density to be
\bea
&&\mathcal{H}^L_k = \sum_\s\left[\left({k^2\over8m_\s v_\s^2}+{\mu_\s\over 2v_\s^2}\right)|\p_\s|^2+{v_\s^2 k^2\over 2 m_\s}|\theta_\s|^2\right]\nonumber\\
&&\hspace{0.8cm}+{1\over 2k^2}\left|\sum_\s q_j\p_\s \right|^{2}.
\label{Ham}\eea
Note that as expected it is the charges $q_\s$ that couple the different species to one another, as seen in the final term.

\section{Ground State Wave Function}\label{Ground}

For the sake of simplicity, let us now specialize to the case of two species $\s=1,2$. We can readily write the above Hamiltonian density in matrix notation, by defining the following vector fields
\be
\vec\theta_k=\left(\begin{array}{c}\theta_1\\ \theta_2\end{array}\right),\,\,\,\,\,\,
\vec\p_k=\left(\begin{array}{c}\p_1\\ \p_2\end{array}\right),
\ee
and the following matrices
\bea
&&\kk_k = \left(\begin{array}{cc}
{q_1^2\over k^2}+{\mu_1\over v_1^2}+{k^2\over 4 m_1 v_1^2} & {q_1 q_2\over k^2}\\ 
{q_1 q_2\over k^2} & {q_2^2\over k^2}+{\mu_2\over v_2^2}+{k^2\over 4 m_2 v_2^2}
\end{array}\right),\\
&&\ff_k = \left(\begin{array}{cc}
{v_1^2k^2\over m_1} & 0\\ 
0 & {v_2^2k^2\over m_2}
\end{array}\right).
\eea
 This gives the following Hamiltonian density
 \bea
\mathcal{H}^L_k = {1\over2}\vec\p_k^* \,\kk_k\,\vec\p_k + {1\over2}\vec\theta_k^*\, \ff_k\,\vec\theta_k.
\eea

We are now in a position to construct the ground state wave function. Recall that for a single harmonic oscillator with Hamiltonian $H=\kk p^2/2+\ff x^2/2$ the ground state wave function in position space is well known to be $\wf(x)\propto \mbox{exp}(-\sqrt{\ff/\kk}\, x^2/2)$. For the above Hamiltonian $H^L$ we need to generalize this to take into account the non-trivial matrix structure. Some matrix algebra reveals that the result in the field basis is
\be
\wf(\theta_\s)\propto \mbox{exp}\left[-{1\over2}\int\!\!{d^3k\over(2\pi)^3}\,\vec\theta^*_k\,\K_k\,\vec\theta_k\right],
\label{WF}\ee
where $\K_k$ is the following matrix
\be
\K_k = \kk_k^{-1\over4}\ff_k^{1\over2}\kk_k^{-1\over4}.
\ee
Let us now examine this result in some important limits.

\subsection{Superfluid} 
 
Firstly, consider the simple case in which the species are neutral $q_\s=0$. In this case the matrices becomes diagonal, the modes decouple, and we have a set of superfluids. The argument of the exponential in the wave function simplifies into the following form
\be
\vec\theta^*_k\,\K_k\,\vec\theta_k = \sum_\s{v_\s^2\over m_\s}{k^2\over\omega_{\s k}}|\theta_\s|^2,
\label{GroundSuperfluid}\ee
where
\be
\omega_{\s k} = \sqrt{ {\mu_\s k^2\over m_\s}+{k^4\over 4 m_\s^2}}
\ee
is the usual dispersion relation in a superfluid for each species. For long wavelength modes the effective sound speed $c_{\s}$ is 
\be
c_{\s}=\sqrt{\mu_\s\over m_\s}.
\ee
Note that for small $k$, the pre-factor of $|\theta_\s|^2$ in Eq.~(\ref{GroundSuperfluid}) is {\em linear} in $k$ (since $\omega_{\s k}$ is itself linear in $k$)
\be
\vec\theta^*_k\,\K_k\,\vec\theta_k = \sum_\s{v_\s^2\over \sqrt{\mu_\s m_\s}}\,k\,|\theta_\s|^2+\mathcal{O}(k^3).
\label{SFwf}\ee
This will be very important when we come to compute the wave function overlap in the next Section.

\subsection{Superconductor}

Our main interest is the case in which the species are charged $q_\s\neq 0$ and we are studying a superconductor. In this case the full wave function in Eq.~(\ref{WF}) is somewhat complicated. However, of most interest will be the long wavelength behavior, as this will control the overlap of any pair of ground states, as we detail in the next Section. For this we can take $q_\s\neq0$ and then perform a small $k$ expansion in the exponent of the wave function (this can also be viewed as a large $q_\s$ expansion). We expand the above matrix $\K_k$ to the first several leading terms and obtain
\be
\K_k=\left(\begin{array}{cc} 
\ca_{22} k + \cb_{12} k^{3\over2} + \cc_{11} k^2 & -\ca_{12} k + \tilde\cb k^{3\over2}+\cc_{12} k^2 \\
-\ca_{12} k + \tilde\cb k^{3\over2}+\cc_{12} k^2 & \ca_{11} k - \cb_{12} k^{3\over2} + \cc_{22} k^2
\end{array}\right)
\label{SCM}\ee
plus corrections that are $\mathcal{O}(k^3)$. The coefficients $\ca_{ij}, \cb_{ij}, \tilde\cb, \cc_{ij}$ are defined as $\ca_{ij}\equiv q_iq_j\ca, \,\cb_{ij}\equiv q_iq_j\cb,\, \tilde{\cb}\equiv{\cb\over2}(q_2^2-q_1^2),\, \cc_{ij}\equiv q_iq_j\cc$,  where
\bea
&&\ca \equiv {v_1 v_2(\sqrt{m_2}\,q_2^2 v_1+\sqrt{m_1}\,q_1^2 v_2)\over(q_1^2+q_2^2)^{3\over2}\sqrt{m_1m_2(q_2^2v_2^2\mu_1+q_1^2v_1^2\mu_2)}},\\
&&\cb \equiv {2 q_1 q_2(v_1^{3\over2}\sqrt{m_2v_2}-v_2^{3\over2}\sqrt{m_1 v_1})\over(q_1^2+q_2^2)^2\sqrt{m_1m_2}(q_2^2v_2^2\mu_1+q_1^2v_1^2\mu_2)^{1\over4}},\\
&&\cc \equiv {\sqrt{m_2}\,q_1^2v_1+\sqrt{m_1}\,q_2^2v_2\over\sqrt{m_1m_2}(q_1^2+q_2^2)^{5\over2}}.
\eea
If we then expand out the matrix structure that appears in the argument of the exponent of $\wf$ we find
\bea
&&\vec\theta^*_k\,\K_k\,\vec\theta_k = \ca k (q_2\theta_1-q_1\theta_2)^2+\cb k^{3\over2}(q_2\theta_1-q_1\theta_2)(q_1\theta_1+q_2\theta_2) \nonumber\\
&&\hspace{1.5cm}+\cc k^2 (q_1\theta_1+q_2\theta_2)^2+\mathcal{O}(k^3).
\label{ExpSC}\eea
Note the important phase parameter dependencies here: The first term $\sim k$ has dependence on $q_2\theta_1-q_1\theta_2$, while the last term $\sim k^2$ has dependence on $q_1\theta_1+q_2\theta_2$, while the second term $\sim k^{3\over2}$ depends on both. Recall that the $U(1)_{em}$ phase rotations transformations are $\theta_\s({\bf x})\to\theta_\s({\bf x})+q_\s\epsilon$; this evidently does {\em not} affect the first term, but only the final terms.

\section{Wave Function Overlap}\label{Overlap}

We now come to the main issue of comparing the set of ground state wave functions that differ by the symmetry transformations
\be
\theta_\s({\bf x})\to\theta_\s({\bf x})+\alpha_\s
\ee
for different choices of phase rotations $\alpha_\s$. Note that if we choose $\alpha_\s=q_\s\alpha({\bf x})$, with $\alpha(x)\to0$ at infinity, then this represents a (small) gauge transformation and the above wave function, like all wave functions, is in fact gauge invariant; this can be made manifest by simply reinstating $\theta_\s({\bf x})\to\theta_\s({\bf x}) - q_\s\nabla\cdot{\bf A}^L/\nabla^2$, which is a well defined operation for (small) gauge transformations. As emphasized earlier in the paper, the interesting issue is that of {\em global} transformations, with $\alpha_\s$ constant.

However, performing constant phase rotations is awkward in $k$-space, since it would formally involve shifting $\theta_{k}$ by a delta-function $\theta_{\s k}\to\theta_{\s k}+\alpha_\s(2\pi)^3\delta^3({\bf k})$. This means we then need to deal with factors of $k$ multiplying delta-functions. So as $k\to0$, this means we formally have zero times infinity. Since delta functions are a kind of distribution and all distributions are defined in terms of their Fourier transform, then, as we did in Ref.~\cite{Hertzberg:2018kyi}, it is much more transparent to pass to position space. 
Since we are integrating over all of space $\int d^3x$, it will be useful to be precise; mathematically an integral over an infinite domain is of course defined as $\int d^3x=\limV\int_Vd^3x$, i.e., as a limit over a volume $V$ as the volume is taken to infinity. In position space, we will then {\em define} our ground state wave function by its kernel representation
\be
\wf(\theta_\s)\propto\limV\mbox{exp}\left[-{1\over2}\int_V\! d^3x\int_{\V}\! d^3x'\,\vec\theta({\bf x})\K_\uv({\bf x}-{\bf x}')\vec\theta({\bf x}')\right],
\ee
(where $\V$ is the domain that ${\bf x}'$ is integrated over, with volume $V$ centered at ${\bf x}$; again all understood in the limiting sense of $\limV$). Here $\K_\uv$ is a matrix of kernels, defined by
\be
\K_\uv(r) = \int\!{d^3k\over(2\pi)^3}\,M_k\,e^{-i{\bf k}\cdot{\bf r}}\,e^{-k\,\uv}.
\ee
Here $\uv$ is a UV regulator. It will be convenient to first regulate the UV modes, then send $\uv\to0$ limit at the end of the calculation. This makes good physical sense, since the physics associated with the structure of ground state is the infrared behavior of the theory and should not be sensitive to the UV. While there may be other prescriptions to define the ground state, which is subtle due to the fact that one is in infinite volume, what is important is that the above definition is quite natural and we will use it self-consistently throughout this paper. In particular, we will use it for both the superfluid and superconductor cases, and we shall see that it allows the distinct realizations of its symmetry structure to be made manifest. While other prescriptions can hide these important distinctions and so are less useful for our purposes. 

In position space we also note that the the complete wave function should be periodic under $\theta_\s\to\theta_\s+2\pi n_\s$, where $n_\s$ is an integer. This is easily ensured by defining the improved wave function as
\be
\tilde\psi(\theta_\s)\propto\sum_{n_1,n_2}\psi(\theta_1+2\pi n_1,\theta_2+2\pi n_2),
\ee
and furthermore, the final result is to be normalized appropriately.

Let us now consider a pair of ground state wave functions: One of them, $|G\rangle$, centered around $\theta_\s=0$ and the other one, $|G'\rangle$, centered around $\theta_\s=\alpha_\s$. The (normalized) overlap between these two wave functions is given by the integral
\be
\langle G|G'\rangle = {\int\mathcal{D}\theta_1\!\mathcal{D}\theta_2\,\tilde\wf(\theta_\s)\tilde\wf(\theta_\s+\alpha_\s)\over
\int\mathcal{D}\theta_1\!\mathcal{D}\theta_2\,\tilde\wf(\theta_\s)\tilde\wf(\theta_\s)}.
\ee
We can readily compute this integral as it is Gaussian. Since $\K_\uv({\bf x}-{\bf x}')$ is only a function of the difference between the position vectors, we can easily perform one of the spatial integrals to obtain a volume factor $V$, leaving a single spatial integral left over. We obtain
\be
\langle G|G'\rangle 
\propto\limV\sum_{n_1,n_2}\mbox{exp}\left[-{V\over4}\vec{\alpha}_{n_1n_2}\int_V d^3 r\, \K_\uv(r)\,\vec\alpha_{n_1n2}\right],
\ee
where we defined a vector of constant phase rotations
\be
\vec\alpha_{n1n2}=\left(\begin{array}{c}\alpha_1+2\pi n_1\\ \alpha_2+2\pi n_2\end{array}\right).
\ee
The above integral $\int d^3r\,\K_\uv(r)$ can be more readily understood by re-writing $M_\uv(r)$ in terms of the Laplacian of another matrix of kernels $\Kr_\uv(r)$ defined implicitly by
\be
\K_\uv(r)=-\nabla^2\Kr_\uv(r).
\ee
In terms of a Fourier transform we can define this by
\be
\Kr_\uv(r) = \int\!{d^3k\over(2\pi)^3}\,{1\over k^2}\,\K_k\,e^{-i{\bf k}\cdot{\bf r}}\,e^{-k\,\uv},
\label{Jdef}\ee
where we inserted an extra factor of $1/k^2$ in the integrand (and we used the fact that $\uv$ is very small).

Using the divergence theorem, the wave function overlap can then be given by the following boundary term
\be
\langle \g|\g'\rangle 
\propto\limV\sum_{n_1,n_2}\mbox{exp}\left[{V\over4}\vec{\alpha}_{n_1n_2}\oint d{\bf S}\cdot\nabla\Kr_\uv(r)\,\vec\alpha_{n_1n2}\right],
\label{OverlapJ}\ee
where $d{\bf S}$ is an infinitesimal surface area vector that points radially outward, which bounds the domain of integration $V$. In this representation it is now clear that the UV has decoupled, as the boundary term is purely an IR effect. In other words, we can now send $\uv\to0$ to evaluate the above kernels, since we know that we do not need to evaluate $\Kr(r)$ as $r\to 0$; we only need to evaluate $\Kr(r)$ at large $r$.

\subsection{Superfluid}

In the case of the superfluid, recall that the matrix of kernels $\K_k$ is diagonal, and hence the $\Kr(r)$ will be diagonal too. Using the leading order result from Eq.~(\ref{SFwf}), in which $\K_k\sim k$, we have
\be
J(r)_{ij} = \delta_{ij}\,{v_\s^2\over\sqrt{\mu_\s m_\s}} f_1(r).
\ee
Here we have defined the function $f_1(r)$, which is a special case of the Fourier transform of inverse powers of $k$, defined by
\be
f_p(r) \equiv \lim_{\uv\to 0} \int\!{d^3k\over(2\pi)^3}{1\over k^p}e^{-i{\bf k}\cdot{\bf r}}\,e^{-k\uv}.
\ee
For the special case of $p=1$, it is readily found to be
\be
f_1(r)={1\over 2\pi^2 r^2}.
\label{f1}\ee
Note that this has significant support at large $r$.  By taking the gradient of $\Kr(r)$, inserting into Eq.~(\ref{OverlapJ}), and taking the boundary to be sphere of radius $R$, we obtain the following result for the overlap
\be
\langle \g|\g'\rangle =\limR{1\over\mathcal{N}}\sum_{n_1n_2}\mbox{exp}\left[-\sum_\s{v_\s^2\over\sqrt{\mu_\s m_\s}}{4\over3}R^2(\alpha_\s+2\pi n_\s)^2\right]\!,
\ee
where the normalization factor $\mathcal{N}$ is simply equal to the numerator with $\alpha_\s=0$. For any finite $\alpha_\s$ in the domain $0<\alpha_\s<2\pi$ we can readily approximate the sum over $n_1,n_2$ by just the $n_1=n_2=0$ terms. This gives the simpler expression
\be
\langle \g|\g'\rangle =\limR\mbox{exp}\left[-\sum_\s{v_\s^2\over\sqrt{\mu_\s m_\s}}{4\over3}R^2\alpha_\s^2\right].
\ee
Evidently, for $\alpha_j\neq0$, the wave function overlap falls off exponentially to $\langle \g|\g'\rangle=0$ by taking the $R\to\infty$ limit, and this occurs for each of the independent modes. This is connected to the usual notion of SSB of a global symmetry and the fact that each mode is associated with its own independent Goldstone (phonon).

\subsection{Superconductor}

In the case of the superconductor with non-zero charges, we return to our expression in Eq.~(\ref{SCM}) for the leading IR contribution to the kernel $\K_k$. In this case the leading dependence for small $k$ include $k$, $k^{3/2}$, and $k^2$. Which of these dominates will depend on the particular choice of phase rotations, as seen in Eq.~(\ref{ExpSC}). To compute the various contributions to $\Kr(r)$, we therefore need to divide by a factor of $k^2$ (recall Eq.~(\ref{Jdef}) compute the Fourier transform of $1/k$, $1/\sqrt{k}$, and $1$. The Fourier transform of $1/k$ is denoted $f_1(r)$ and was reported earlier in Eq.~(\ref{f1}); it has significant support at large $r$ and scales as $\sim1/r^2$. Similarly the Fourier transform of $1/\sqrt{k}$ is
\be
f_{1\over2}(r) = {1\over 4\sqrt{2}\,\pi^{3/2}r^{5/2}},
\ee
which also has somewhat significant support at large $r$. On the other hand, the Fourier transform of $1$ is known to be just a delta-function
\be
f_0(r) = \delta^3({\bf r}),
\ee
and has no support at all at large $r$. Hence the terms in $\K_k$ that involve $k^2$ do not contribute at all to the wave function overlap at large volume. In fact these are precisely the terms that arise from electric transformations $\alpha_\s=q_\s\epsilon$, which is only non-zero for the $k^2$ terms (as well as higher order terms, that all involve even powers of $k$; all are associated with delta-functions and do not contribute at large volume to the overlap).

Using these results, we find that the matrix $\Kr(r)$ for non-zero $r$ is given by
\be
\Kr(r)=\left(\begin{array}{cc} 
{\ca_{22}\over2\pi^2 r^2} + {\cb_{12}\over4\sqrt{2}\,\pi^{3/2}r^{5/2}} & -{\ca_{12}\over2\pi^2 r^2} + {\tilde\cb\over4\sqrt{2}\,\pi^{3/2}r^{5/2}} \\
-{\ca_{12}\over2\pi^2 r^2} + {\tilde\cb\over4\sqrt{2}\,\pi^{3/2}r^{5/2}} & {\ca_{11}\over2\pi^2 r^2} - {\cb_{12}\over4\sqrt{2}\,\pi^{3/2}r^{5/2}}
\end{array}\right)
\label{SCJ}\ee
Inserting this into the general expression for the wave function overlap Eq.~(\ref{OverlapJ}) and again evaluating the integral on a sphere of radius $R$, we obtain our primary result
\bea
&&\langle \g|\g'\rangle = \limR{1\over\mathcal{N}}\sum_{n_1n_2}\mbox{exp}\Big{[}-\ca{4\over3}R^2(q_2\tilde\alpha_1-q_1\tilde\alpha_2)^2\nonumber\\
&&\hspace{2cm}-\cb{5\sqrt{\pi}\over8\sqrt{2}}R^{3/2}(q_2\tilde\alpha_1-q_1\tilde\alpha_2)(q_1\tilde\alpha_1+q_2\tilde\alpha_2) \Big{]}\,\,\,\,\,\,\,\,\,\,\,\,
\label{PR}\eea
where $\tilde\alpha_\s\equiv\alpha_\s+2\pi n_\s$. As before, the normalization factor is simply equal to the numerator with $\alpha_\s=0$. 

For a typical choice of phase rotations $\alpha_\s$, the argument of the exponent is non-zero. This leads to the wave function overlap $\to0$ exponentially fast zero by taking the $R\to\infty$ limit. For $0<\alpha_\s<2\pi$ and for $q_2\alpha_1-q_1\alpha_2\neq 0$ the leading fall off is provided by the first term in the exponent and for $n_1=n_2=0$, giving the leading fall off of the wave function as
\be
\langle \g|\g'\rangle = \limR\mbox{exp}\left[-\ca{4\over3}R^2(q_2\alpha_1-q_1\alpha_2)^2\right].
\ee
So for generic $\alpha_\s$ this is once again zero, as we saw in the above superfluid case; indeed it is displaying the usual notion of SSB and there is a corresponding Goldstone mode, which we will describe in more detail in the next Section.

However, notice that there is one, and only one, special choice of phase rotations that does {\em not} leads to $\langle \g|\g'\rangle=1$; namely if we perform a regular $U(1)_{em}$ phase rotation
\be
\alpha_\s = q_\s\,\epsilon,
\ee
(where $\epsilon$ is a common factor). This is the one special combination that sets
\be
q_2\alpha_1 - q_1\alpha_2 = 0,
\ee
leading towards vanishment of both terms in the argument of the exponent in the wave function overlap eq.~(\ref{PR}). In fact we have checked that it vanishes for all higher order contributions to the wave function too; by including all higher order terms, one finds that the delta-function for $J$ is smeared out into a function that is exponentially suppressed at large distances (this has a physical reason connected to the fact that the photon is massive in this phase, as we will return to in Section \ref{CDF}), and so does not contribute as $R\to\infty$.  Hence for this particular transformation, we have $|G'\rangle=|G\rangle$. However, any other transformation we have $\langle G|G'\rangle=0$; so these accompanying transformations act as a type of ``custodial" symmetry that is spontaneously broken in the usual sense. The fact that for the regular $U(1)_{em}$ transformation, the behavior is qualitatively different, has a physical explanation: In physical terms it is closely connected to the Goldstone boson structure, i.e., the fact that this Goldstone is removed, leading to gapped plasma oscillations instead. Furthermore, it is connected to the exponential fall off of the electric field, causing the charges to go to zero, as we discuss shortly.

\section{Conserved Quantities}\label{Conserved}

To understand this result further, let us examine the possible conserved quantities in the system. Naively there is a conserved quantity for each rotation $\alpha_\s$. Indeed this is true when expanding around the vacuum. However, when expanding around the superconducting condensate, it is more subtle.

\subsection{Total Charge}

Recall that each species has a corresponding particle number given by 
\be
N_\s = \int d^3x|\Psi_\s({\bf x},t)|^2.
\ee
In this section we will study the classical field evolution for simplicity. The leading order fluctuations in particle number $\Delta N_\s$ around the homogeneous background are given by
\be
\Delta N_\s = 2 v_\s \int d^3 x\,\m_\s({\bf x},t) = 2 v_\s\,\m_\s|_{k\to0},
\ee
where in the second step we have expressed the spatial integral as the zero mode of the Fourier space representation. The second time derivative of this is
\be
\Delta \ddot{N}_\s = 2 v_\s\,\ddot{\m}_\s|_{k\to0} = {v_\s^2k^2\over m_\s}\dot{\theta}_\s|_{k\to0},
\ee
where in the second step we used the classical equation of motion for $\m_\s$ that follows from the Hamiltonian Eq.~(\ref{Ham}).  This can be determined to be
\be
\Delta\ddot{N}_\s = -{v_\s^2q_\s\over m_\s}Q,
\ee
where we used the classical equation of motion for $\theta_\s$. Here the total electric charge $Q$ is
\be
Q=\sum_\s q_\s \,\Delta N_\s.
\ee
In general we see that the particle numbers in a superconductor are typically not conserved if the total integrated charge fluctuation is non-zero (this is to be contrasted to the case of expanding around the vacuum). Related to this, we can compute the time evolution of the charge itself. The last 2 equations give
\be
\ddot{Q}=-\sum_j{v_\s^2 q_\s^2\over m_\s}\,Q.
\label{Qevolution}\ee
Hence the electric charge is not conserved, but oscillates in any enclosed region if its total initial value is non-zero; these are the familiar plasma oscillations.

\subsection{Charge Density Fluctuations and Screened Electric Field}\label{CDF}

As detailed in the appendix (where we also include leading order relativistic corrections for completeness) we can study the charge density fluctuations also. For a multi-fluid system, these fluctuations are complicated, so it will suffice here to report on the case of a single species. One can readily show that in this case the charge density $\rho$ is related to $\dot\theta$ (in Coulomb gauge) by
\be
\rho_k = -{q\,k^2\over{k^4\over 4mv^2}+{\mu\,k^2\over v^2}+q^2}\dot\theta_k
\label{rhoothet}\ee
Then one can show that the equation of motion for $\theta_k$ is given by
\begin{equation}
\ddot{\theta}_k = -\left[m_{\tiny\mbox{eff}}^2 + \dfrac{k^2}{4m^2}\left(k^2 + m_{\xi}^2\right)\right]\theta_k
\label{eq:equation.of.motion.k}
\end{equation}
where $m_{\tiny\mbox{eff}}= v^2q^2/m$ and $m_\xi^2=4\mu m$. One can then replace $k^2\to-\nabla^2$ to turn this into a wave equation back in position space. 
By taking a time derivative of this equation, and using Eq.~(\ref{rhoothet}), we see that the exact same equation is obeyed by the charge density itself. By integrating this over space and dropping boundary terms, we obtain the earlier Eq.~(\ref{Qevolution}) (if we simply generalize again to multiple species). 

However, this form of the fluctuations equation, reveals something very important: Suppose we consider initial conditions provided by $\theta$ and $\dot\theta$ that are localized, i.e., they have support in the bulk, but die away rapidly towards the boundary. These are associated with perfectly reasonably initial conditions, with finite energy, etc. Then we see that the total charge is of a special form. Consider the low $k$ limit of Eq.~(\ref{rhoothet})
\be
\rho_k = -{k^2\over q}\dot\theta_k\,\,\,\,\,\,\,\,\,(\mbox{small}\,\,\,k)
\label{rhoothet2}\ee
If we then write out the expression for the total integrated charge, it is
\bea
Q&=&\limV\int_V\! d^3x\,\rho({\bf x},t) = -{1\over q}\limV\int_V\! d^3x\,\nabla^2\dot\theta\\
&=&-{1\over q} \limS\oint d\bf{S}\cdot\nabla\dot\theta
\eea
where $\oint dS$ indicates the integral over a boundary surface, taken in the limit out towards infinity, by use of the divergence theorem.  For a localized initial condition, i.e., $\dot\theta\to 0$ at spatial infinity, then we have that the enclosed charge $Q\to 0$. Hence this gives a conserved charge, but only in a trivial sense, i.e., $Q=0$ and it remains 0. On the other hand, if one has a charge density that extends all the way out towards infinity, then one can have a non-zero $Q$, but it will no longer be conserved, as in Eq.~(\ref{Qevolution}). 

This phenomenon is closed related to the screening of the electric field in the superconductor. For any number of fields, and using Gauss' law, we have
\be
\limS\oint d{\bf S}\cdot{\bf E} = Q
\ee
As we showed above, for local $\theta$ and $\dot\theta$, the integrated $Q$ vanishes. This is connected to the electric field being exponentially suppressed at large distances, which ensures a surface integral over it vanishes at large distance. Hence we recover the idea that the electric field is {\em screened} in a superconductor.  In fact in Coulomb gauge, the Coulomb potential $\phi$ can be shown to also obey the same equation as Eq.~(\ref{eq:equation.of.motion.k}). 
So we see that if $m_\xi \gg \sqrt{m\, m_{\tiny\mbox{eff}}}$, the electric field is screened over lengths $\sim m_\xi/(m\,m_{\tiny\mbox{eff}})$ to leading order, whereas if $m_\xi \gg \sqrt{m\, m_{\tiny\mbox{eff}}}$, it is screened over lengths $\sim 1/\sqrt{m\, m_{\tiny\mbox{eff}}}$. In either of the cases, we see that it is shielded even more strongly than the magnetic field (which is screened over lengths $\sim 1/m_{\tiny\mbox{eff}}$).

Since the charge integrates to $Q=0$ (for localized sources), this has ramifications for the properties of the vacuum. In the quantum theory, it will annihilate the vacuum $\hat{Q}|G\rangle=0$.  Since, in the quantum theory, charge is the generator of a symmetry $\hat{S}=e^{i\hat{Q}}$, the corresponding symmetry transformation to another state
\be
|G'\rangle = \hat{S}|G\rangle = e^{i\hat{Q}}|G\rangle=|G\rangle
\ee
is the identity operator, which maps ground states into themselves. This is to be contrasted to other conserved quantities and symmetries that we discuss in the next subsection. 

Furthermore, this is closely connected to the spectrum of particles. In the superfluid case, there are Goldstones. So we can move from one vacuum to another by adding more and more zero-momentum Goldstones, without raising the energy. This is a reflection of the usual notion of multiple distinct ground states. In the superconductor case, the corresponding would-be Goldstone (see others in the next subsection) has become massive (plasma oscillations), so if we start in a vacuum and now add more and more would-be Goldstones, we raise the energy, moving out of the vacuum.

\subsection{Other Combinations}

For a two species system, there is one linear combination of the $\Delta N_\s$ that is conserved. The enclosed perturbation in total mass is given by
\be
\Delta M = \sum m_\s\,\Delta N_\s.
\ee
This evolves according to 
\be
\Delta\ddot{M} = -\sum_\s v_\s^2 q_\s\,Q = 0,
\ee
where in the last step we used the condition that the background total charge density $\rho_0= \sum_\s v_\s^2 q_\s=0$, so that we are expanding around a neutral superconductor (to be clear, one needs to extend the sum over $j$ to include the positively charged heavy nuclei, as they cannot be ignored when we multiply throughout by the mass of the particles). Hence there is a single conserved quantity associated with these internal symmetries; which is the conservation of mass. Its corresponding Goldstone is a phonon. The associated phase rotation transformations that are generated by this conserved quantity are
\be
\theta_\s({\bf x}) \to \theta_\s({\bf x}) + m_\s \epsilon.
\ee
This is a symmetry transformation that is spontaneously broken by the ground state in the usual way. In this particular construction, it is acting as a kind of ``custodial" symmetry that is spontaneously broken in the super-conductor phase, while it remains unbroken in the regular phase. In fact the related Galilean symmetry of boosts is also spontaneously broken. We see that this behaves  differently to the $U(1)_{em}$ ($\theta_\s({\bf x})\to\theta_\s({\bf x})+q_\s\epsilon$); in this case, there is no regular conserved charge, so this symmetry can be interpreted as being {\em removed} in this phase.

\section{Conclusions}\label{Conclusions}

In this work we have explicitly computed the wave function overlap between ground states, defined in a specific prescription through the position space kernel representation, in nonrelativistic systems of condensed bosons, either modeling superconductivity or superfluidity. 
We showed that while a generic phase rotation transformation of the nonrelativistic Schrodinger field does indeed lead to vanishing overlap for any phase rotations, associated with those of a superfluid. For a super-conductor, there is a special combination of phase rotations that does not lead to a new state which are the standard form of electromagnetic phase rotations. In the literature, this is often summarized by the language that for ``gauge symmetry breaking, the apparently different ground states are related by gauge transformations" in the bulk of a superconductor.\footnote{Other related comments appear in the literature, including Ref.~\cite{Poniatowski}, which claims that in the bulk of a superconductor all the apparently distinct states are in fact the same physical state; to be contrasted to the superfluid case.}
However, here we wish to move beyond language and emphasize that there is a very physical phenomenon occurring: the electric field gets screened; this leads to the charge $Q=\limS\oint d{\bf S}\cdot{\bf E}$ formally vanishing, and so as an operator, it maps a ground state into itself. Moreover, in the superfluid case, one can add a large number of massless Goldstones of zero momenta to generate a new $|G'\rangle$, while in the superconductor case, the spectrum is gapped, so adding massive photons would raise the system out of the ground state.

Note that this in no way undermines the idea of a {\em phase transition} from the normal phase to the superconductor. One can define order parameters, such as $\langle|V^{-1}\!\int_V d^3x\,\hat{\Psi}_i|\rangle$ (note the absolute value; it is unchanged under a phase rotation transformation), which are non-zero in the superconducting phase, while vanishing in the regular phase. This connects to the more familiar notion of SSB in field theories. Furthermore, one can express this in entirely particle physics terms: the former phase has a massless photon with long-ranged electric/magnetic fields, while the latter phase has a massive photon with short ranged electric/magnetic fields.

It is important to note that when the phase transition to the super-conducting state occurs, our ground state overlap exhibits a standard notion of SSB of related global symmetries. In this context of multiple bosonic fields, another is associated with mass conservation with phonons acting as the Goldstones, among other possibilities depending on the number of fields. These additional symmetries that can undergo the standard form of global SSB can act as their own diagnostic to determine the physical phase that the theory is in (Higgs or Coulomb, etc) since the phase transition is accompanied by their breaking; so they act as a kind of custodial symmetry breaking (related ideas appear in Ref.~\cite{Matsuyama:2019lei}). 

Our primary result is that our overlap furnishes a clear mapping to the Goldstone structure. In particular, with two-bosonic fields: if neither phase rotation exhibits $\langle G|G'\rangle=0$, we are in a normal phase; if one combination of the phase rotations exhibits$\langle G|G'\rangle=0$, we are in a superconducting phase; if both phase rotations exhibits $\langle G|G'\rangle=0$, we are in a superfluid phase. We noted that in the case of the super-conductor, the different structure in the overlap is related to the fact that the electric field is {\em screened} in this phase. So by using Gauss' law the corresponding charge, expressed as a boundary term, vanishing. Since it acts as a symmetry operator in the quantum theory, it maps a ground state into itself. However, for the superfluid, its conserved ``charges" (or particle numbers) cannot be expressed in terms of vanishing boundary terms; they are non-vanishing and drive the overlap to vanish for any phase rotation. So this is all in accord with good physical principles. 

Our focus here has only been on the bulk and the direct computation of the overlap wave function, which we believe to be a new result. This construction is useful because it is nicely in one-to-one correspondence with the presence, or lack thereof, of Goldstone modes. One of course should expect that the existence of Goldstones should be reflected in the structure of the ground state wave function. Our prescription to define the ground state at infinite volume through its position space kernel representation, has led to the Goldstone bosons being directly connected to whether the overlap $\langle G|G'\rangle=0$ or not. This is an especially useful diagnostic for the physical phase of the system as it is gauge invariant (unlike most other order parameters), and may be applied in more general settings with richer dynamics. 

On the other hand, if one has multiple finite size superconductors, there can be interesting boundary effects, including the Josephson effect \cite{Josephson1,Josephson2}, when multiple superconductors are brought in contact with each other. The system can release some energy, in the form of the Josephson current, to relax to an even lower energy state. This provides other important features of the ultimate fate of SSB \cite{Greiter}. However, the focus of this work has been on the wave function in the bulk of a single material and relates directly to the physical phase that the system is in, including its energy spectrum. 

These results are in accord with our earlier work in the Standard Model \cite{Hertzberg:2018kyi}. Other directions to consider are more complicated condensed matter systems, including those that exhibit strong coupling, and various other phases. Further applications may be to other areas in which SSB may play a role, including ideas in particle physics \cite{Georgi:1974sy,Fritzsch:1974nn,Ross} and cosmology \cite{Gripaios:2004ms,Hinterbichler:2011qk,Boyanovsky:2012nd,Hertzberg:2014iza,Klein:2017npd}.

\section*{Acknowledgments}

MPH is supported in part by National Science Foundation grants No. PHY-1720332, PHY-2013953.

\appendix

\section{Relativistic Corrections}\label{RA}

In order to study relativistic effects and the screening of electric field in a superconductor in more detail, in this appendix we consider only one bosonic Schrodinger field $\psi$ (the Cooper pair field), and also include the relativistic term in the Lagrangian \eqref{eqn:Lagrangian} for illustration purposes:
\begin{equation}
\mathcal{L} \rightarrow \mathcal{L} + \delta\dfrac{1}{2m}\left|\dot{\Psi} + iq\phi\psi\right|^2.
\end{equation}
However, as we will see, the final result will be insensitive to this term since we are always interested in the nonrelativistic limit $k/m \ll 1$, and to make this point clear, we have introduced a ``switch" $\delta$ in the front; $\delta=0$ is the nonrelativistic theory, while $\delta=1$ includes the relativistic correction. 

Working in the Coulomb gauge, the leading order longitudinal Lagrangian (c.f. \ref{longitudinal}) for the fluctuations (c.f. \ref{eqn:perturbed.field}), after solving for the scalar potential $\phi$ and inserting it back as before, is therefore
\begin{eqnarray}
\mathcal{L}_L &=& -v\,\eta^*_{k}\left(\dfrac{k^2}{\omega_{\tiny\mbox{eff}}^2}\right)\dot{\theta}_{k} + c.c - \dfrac{v^2\,k^2}{2m}\left|\theta_{k}\right|^2 - \eta^*_{k}\left(\dfrac{1}{2m}\Lambda^2\right)\eta_{k}\nonumber\\
&+& \delta\dfrac{1}{2m}\left[|\dot{\eta}_{k}|^2 + \dfrac{v^2\,k^2}{\omega_{\tiny\mbox{eff}}^2}\left|\dot{\theta}_{k}\right|^2\right].
\label{eq:weak.field.Lagrangian}
\end{eqnarray}
where
\begin{eqnarray}
\omega_{\tiny\mbox{eff}}^2 &\equiv& k^2 + \delta\dfrac{v^2q^2}{m} \equiv k^2 + m_{\tiny\mbox{eff}}^2\delta,\nonumber\\
\Lambda^2 &\equiv& k^2 + 4\,\mu\,m + \dfrac{4m_{\tiny\mbox{eff}}^2m^2}{\omega_{\tiny\mbox{eff}}^2} \equiv k^2 + m_{\xi}^2 + \dfrac{4m_{\tiny\mbox{eff}}^2m^2}{\omega_{\tiny\mbox{eff}}^2}.
\end{eqnarray}
From this Lagrangian, it is suggestive that the radial degree of freedom $\eta$ is super-massive because of the factor of $4m^2$ in $\Lambda^2$. So in the limit when $k \ll m$, we can neglect the $\dot\eta^2_{k}$ term and then it becomes a constraint variable which we can solve for
\begin{equation}
\eta_{k} \approx -\dfrac{2m\,v\,k^2}{\omega^2_{\tiny\mbox{eff}}\Lambda^2}\dot\theta_k
\end{equation}
and insert back into the Lagrangian to obtain the following Lagrangian for the longitudinal degree of freedom
\begin{eqnarray}
\mathcal{L}_L &\approx& \dfrac{m\,v^2}{2\omega_{\tiny\mbox{eff}}^2}\left(\delta\dfrac{1}{m} + \dfrac{4k^2}{\Lambda^2\omega^2_{\tiny\mbox{eff}}}\right)\left|\dot{\theta}_{k}\right|^2 - \dfrac{v^2k^2}{2m}\left|\theta_{k}\right|^2.
\end{eqnarray}
The equation of motion for this degree of freedom to leading order (the coherence length $\xi$ has to be much larger than the Compton wavelength $l_{\tiny\mbox{cp}}$ of the bosonic degree of freedom, i.e. we can expand in $m_{\xi}/m$) and we recover
\begin{equation}
\ddot{\theta} = -\left[m_{\tiny\mbox{eff}}^2 - \dfrac{\nabla^2}{4m^2}\left(-\nabla^2 + m_{\xi}^2\right)\right]{\theta}.
\label{eq:equation.of.motion.2}
\end{equation}
as we reported earlier when beginning in the exact non-relativistic theory. Note that there is no dependence on $\delta$ here. Now since the Coulomb scalar potential $\phi = \hat{O}\theta$ where $\hat{O}$ is a linear operator, so it too satisfies this.

\end{document}